# Raman Signatures of Broken Inversion Symmetry and In-plane Anisotropy in Type-II Weyl Semimetal Candidate TaIrTe$_4$

Yinan Liu[#,1], Qiangqiang Gu[#,1], Yu Peng[2], Shaomian Qi[1], Na Zhang[3], Yinong Zhang[1], Xiumei Ma[4], Rui Zhu[4], Lianming Tong[3], Ji Feng[*,1,5], Zheng Liu[*,2,6] and Jian-Hao Chen[*,1,5]

[1]International Center for Quantum Materials, School of Physics, Peking University, NO. 5 Yiheyuan Road, Beijing, China, 100871
[2]Centre for Programmed Materials, School of Materials Science and Engineering, Nanyang Technological University, Singapore 639798, Singapore
[3]College of Chemistry and Molecular Engineering, Peking University, NO. 5 Yiheyuan Road, Beijing, China, 100871
[4]Electron Microscopy Laboratory, School of Physics, Peking University, Beijing 100871, China
[5]Collaborative Innovation Center of Quantum Matter, Beijing 100871, China
[6]NOVITAS, Nanoelectronics Centre of Excellence, School of Electrical and Electronic Engineering, Nanyang Technological University, Singapore 639798, Singapore

*Corresponding authors:
Ji Feng (jfeng11@pku.edu.cn), Zheng Liu (z.liu@ntu.edu.sg) and
Jian-Hao Chen (chenjianhao@pku.edu.cn)

# These authors contributed equally to this work

**ABSTRACT**

The layered ternary compound TaIrTe$_4$ is an important candidate to host the recently predicted type-II Weyl Fermions. However, a direct and definitive proof of the absence of inversion symmetry in this material, a prerequisite for the existence of Weyl Fermions, has so far remained evasive. Herein, an unambiguous identification of the broken inversion symmetry in TaIrTe$_4$ is established using angle-resolved polarized Raman spectroscopy. Combining with high-resolution transmission electron microscopy, we demonstrate an efficient and nondestructive recipe to determine the exact crystallographic orientation of TaIrTe$_4$ crystals. Such technique could be extended to the fast identification and characterization of other type-II Weyl Fermions candidates. A surprisingly strong in-plane electrical anisotropy in TaIrTe$_4$ thin flakes is also revealed, up to 200% at 10K, which is the strongest known electrical anisotropy for materials with comparable carrier density, notably in such good metals as copper and silver.

**Introduction**

The recent discovery of Weyl semimetals has attracted substantial attention among materials scientists and condensed matter physicists[1-3]. It was predicted that two types of Weyl fermions[4, 5] may exist in solids with broken spatial inversion symmetry or time-reversal symmetry[2, 6]. In type-I Weyl semimetal the bulk Fermi surface shrinks to a point at the Weyl node with conserved Lorentz invariance[7]; whereas in type-II Weyl semimetal the Weyl points appear at the topologically protected touching points between electron and hole pockets with titled Weyl cones[5].

While considerable progress in both theory and experiment has been made on type-I Weyl semimetals[1, 2, 8], only a few type-II Weyl semimetals have been identified[5, 9, 10]. TaIrTe$_4$ has recently been proposed to be a type-II Weyl semimetal candidate[7]. As a ternary variant of WTe$_2$, TaIrTe$_4$ is a layered material with interestingly strengthened Te-Te bonds and various possible crystal symmetry[11], referred to as the monoclinic (1$T'$ phase) and orthorhombic structures ($T_d$ phase). While the monoclinic 1$T'$ phase has the centrosymmetric space group P2$_1$/$m$, the orthorhombic $T_d$ phase has two possible space groups, the non-centrosymmetric P*mn*2$_1$ and the centrosymmetric P*nmm*[12]. The P*mn*2$_1$ $T_d$ phase TaIrTe$_4$ is predicted to host just four type-II Weyl points, the minimal number of Weyl points hosted by a system with time-reversal invariance[7, 10]. Intensive efforts have been made to search for the related effects of the broken inversion symmetry and the type-II Weyl nodes of the material[2, 6]. However, the definitive signature of the broken spatial inversion symmetry or time-reversal symmetry in TaIrTe$_4$ has yet to be firmly established[13]. Previous X-ray diffraction study was confronted with the uncertainty associated with subtle differences between the P*mn*2$_1$ and the P*nmm* space groups[11]. Recent pump-probe angle-resolved photoemission spectroscopy studies[13, 14] have observed hints of Weyl points and topological Fermi arcs in TaIrTe$_4$ with limited resolution, since the Weyl points and topological Fermi arcs are predicted to reside entirely above the Fermi level in TaIrTe$_4$[13]. For transport studies, a negative longitudinal magnetoresistance, which signifies the chiral anomaly of Weyl Fermions, has not yet been observed in TaIrTe$_4$[15].

In this report, we use angle-resolved polarized Raman spectroscopy, which are directly sensitive to the crystal symmetry, to confirm unambiguously the absence of inversion symmetry in TaIrTe$_4$ bulk crystals as well as in its thin flakes. We observe strong optical and electrical in-plane anisotropy of few-layer TaIrTe$_4$ and provide a rapid and nondestructive method to identify the crystallographic orientation of TaIrTe$_4$. The in-plane electrical anisotropy of TaIrTe$_4$ thin flakes is found to be the strongest among materials with comparably high carrier density, which may enable new architectures in electrical interconnects in future integrated circuits. The results will prove valuable not only for further research into the physics and application of TaIrTe$_4$ but also for exploration of the family of type-II Weyl semimetals.

**Results and discussion**

**Raman spectra of TaIrTe$_4$ with different thickness**

**Figure 1**a represents the side view of the TaIrTe$_4$ lattice structure. Alternating Ta-Ir connections stretch along the crystalline *a*-axis as zigzag atomic chains. These chains hybridize with each other along the *b* direction to form a conducting *ab* plane. A transmission electron microscopy (TEM) image of a TaIrTe$_4$ flake is shown in Figure 1b. Inset of Figure 1b is an atomically resolved TEM image of ultra-thin TaIrTe$_4$, which shows clear orthogonal lattice fringes. The lattice constants of TaIrTe$_4$ can therefore be derived from this image: *a* = 0.3895 nm and *b* = 1.2428 nm, consistent with the results in the literature[11].

Few-layer TaIrTe$_4$ flakes were exfoliated mechanically and transferred onto a silicon substrate with 300 nm SiO$_2$. The flake thickness measured by atomic force microscopy (AFM) is shown in Figure 1c. Having a layer-to-layer spacing of 1.32 nm[11], we can clearly identify TaIrTe$_4$ with thicknesses down to around 1.5 nm, corresponding to single unit cell of TaIrTe$_4$. In this work, we focus on flakes with ten layers or more with large enough size to perform optical and electrical measurements.

Figure 1d shows the thickness dependent Raman spectra of TaIrTe$_4$ flakes. It can be seen that the Raman intensity of thin flakes (< 20 nm) is relatively strong compared to the thick flakes (> 20 nm). We attribute this phenomenon to the interference enhancement effect[16]. Contrary to other layered materials, the position of the Raman peaks of TaIrTe$_4$ samples at different thicknesses did not change significantly[17]. Throughout the rest of this study, we shall focus on the flakes with thickness less than 20 nm.

**Raman signature of broken inversion symmetry in TaIrTe$_4$**

**Figure 2**a shows polarized Raman spectra (solid black curve) collected in a parallel polarization configuration from a 19nm sample, where both the incident and scattered light polarize along the crystallographic *a*-axis. We shall focus on symmetry analysis here and discuss optical anisotropy as well as crystallographic directions of TaIrTe$_4$ in the next section.

Among the possible structures of TaIrTe$_4$, the only non-centrosymmetric structure belongs to the space group *Pmn*2$_1$ (point group $C_{2v}$). Thus, we first analyze the irreducible representations of group $C_{2v}$ and make comparison with experimental data. The unit cell of TaIrTe$_4$ consists of 24 atoms, which support 72 phonon modes. Given the point group symmetry of the crystal, 69 optical phonons are Raman-active, which include $23A_1+12A_2+11B_1+23B_2$, where $A_{1,2}$ and $B_{1,2}$ are the irreducible representations of the point group $C_{2v}$.

For the given experimental configuration used to collect data in Figure 2a, only $A_1$ phonons are Raman active. The frequencies of all the $A_1$ phonon were determined by first-principles calculations and were indicated by red and blue vertical bars under the experimental Raman spectra in Figure 2a. It is evident that the frequencies of the experimental Raman peaks agree well with the calculations, with an uncertainty less than 0.46 meV in phonon energy.

In order to establish a direct connection between specific Raman signal and the broken inversion symmetry in the crystal, we invoke an analysis assuming the **existence** of inversion center located in the $TaIrTe_4$ atomic layer between two Ir atoms (see supplementary information section 1 for details). In this case, the point group of the crystal would become $D_{2h}$ and the $A_1$ irreducible representation in point group $C_{2v}$ splits into the parity-even $A_g$ and the parity-odd $B_{1u}$ representations in $D_{2h}$. The $B_{1u}$ phonons are Raman inactive and thus should not be detected by Raman measurements. However, within the experimentally explored energy window, we have identified eight experimental Raman peaks that would belong to the $B_{1u}$ irreducible representation in $D_{2h}$. These eight phonons are highlighted with red vertical bars in Figure 2a labeled by capital letters A-H, and their vibrational patterns are shown in Figure 2b. Note that some of these phonons (especially mode D) have strong Raman signals, indicating a strong inversion symmetry breaking, consistent with previously proposed, large Weyl node separation in $TaIrTe_4$[7]. Mode D also has angle-dependence that matching perfectly with Raman tensor analysis (see Supplementary information 4), excluding the possibility of defect signals. Thus the observation of mode D in our polarized Raman experiments can be seen as a tell-tale signature of broken inversion symmetry, which makes TaIrTe$_4$ possible to be a type-II Weyl semimetal.

**Optical in-plane anisotropy**

Now we consider angle-resolved Raman spectra of the TaIrTe$_4$ samples. Freshly exfoliated thin TaIrTe$_4$ samples on 300 nm SiO$_2$ were mounted onto a rotation stage. The Raman spectra were taken using a 633 nm He-Ne laser. **Figure 3**a shows an optical image of a TaIrTe$_4$ flake with a thickness around 12 nm. The *x*-axis was assigned to be along the naturally occur crystallographic edge as shown in Figure 3a. Later we will show that this edge indeed corresponds to the *a*-axis (zigzag direction of the Ta-Ir chains) of the TaIrTe$_4$ crystal. Figure 3b illustrates the experimental setup for the parallel-polarized and cross-polarized Raman scattering of the TaIrTe$_4$ samples. Figure 3c plots the angular dependences of the Raman intensity spectra in the parallel-polarized configuration. The results of the cross-polarized configuration are reported in supplementary **Figure S3**. The sample rotation angle $\theta$ is in the range of 0-360°.

**Figure 4** plotted in polar figures the intensity of six major Raman peaks evolving with angles in parallel-polarized configuration. Also shown are the atomic schematics of the corresponding phonon modes. Two types of modes are found in the parallel-polarized configuration. The first type has a 2-lobed shape with two maximum intensities at angles $\theta = 180° - \theta_0$ and $\theta = 360° - \theta_0$ ($\theta_0$ was used to denote the initial angle between the incident laser polarization and the *x*-axis); the second type has a 4-lobed shape with maximum intensities at four angles. In the cross-polarized configuration, all modes yield 4-lobed shapes (Figure S3). Similar Raman spectra from TaIrTe$_4$ of two additional thicknesses are shown in supplementary **Figure S4**. The results show that the anisotropic Raman spectra of TaIrTe$_4$ flakes have no clear thickness dependence for flakes thicker than 12nm.

Quantitative analysis on the observed anisotropic phenomena can be made, based on the group theory, from Raman tensors and density functional theory (DFT) calculations. The intensity of Raman signals from these modes can be expressed as[18]:

$$I \propto |\boldsymbol{e_i} \cdot \boldsymbol{R} \cdot \boldsymbol{e_s}|^2 \qquad (1)$$

where $\boldsymbol{e_i}$ and $\boldsymbol{e_s}$ are the unit vectors of the incident and scattered light's polarization and $\boldsymbol{R}$ is the Raman tensor for a certain vibration mode. The Raman tensors corresponding to the $A_1$, $A_2$, $B_1$ and $B_2$ modes are[19]:

$$R_{A_1} = \begin{pmatrix} d & 0 & 0 \\ 0 & f & 0 \\ 0 & 0 & g \end{pmatrix} \quad R_{A_2} = \begin{pmatrix} 0 & h & 0 \\ h & 0 & 0 \\ 0 & 0 & 0 \end{pmatrix} \quad R_{B_1} = \begin{pmatrix} 0 & 0 & 0 \\ 0 & 0 & k \\ 0 & k & 0 \end{pmatrix} \quad R_{B_2} = \begin{pmatrix} 0 & 0 & l \\ 0 & 0 & 0 \\ l & 0 & 0 \end{pmatrix} (2),$$

where $d$, $f$, $g$, $h$, $k$ and $l$ are the tensor elements determined by the cross section of Raman scattering.

In our experiment, the laser shines perpendicular to the (001) surface and thus for a sample with rotation angle $\theta$, $\boldsymbol{e_i} = (\cos(\theta + \theta_0), \sin(\theta + \theta_0), 0)$ for the incident light, and $\boldsymbol{e_s} = (\cos(\theta + \theta_0), \sin(\theta + \theta_0), 0)$ for the scattered light in the parallel-polarized configuration[20]. According to the backscattering geometry, only $A_1$ and $A_2$ modes have non-zero intensity[20]. The angle-dependent intensity for the $A_1$ and $A_2$ modes can be expressed as:

$$I_{A_1}^{\parallel} \propto d^2\cos^4(\theta + \theta_0) + f^2\sin^4(\theta + \theta_0) + 2df\cos^2(\theta + \theta_0)\sin^2(\theta + \theta_0)\cos 2\delta$$

(3)

$$I_{A_2}^{\parallel} \propto h^2\sin^2 2(\theta + \theta_0) \quad (4)$$

Here $\delta$ is a phase factor which accounts for the light absorption effect on the Raman tensor elements[20, 21] or the birefringence effect[22]. We selected six relatively strong Raman peaks in parallel- and cross-polarized configurations, respectively, to fit to equation (3)-(4). The fitted curves are in good agreement with the experimental data (Figure 4). In particular, possible effects from defects can be excluded from the angular dependence of the major Raman peaks as well as the stability of these peaks under ambient conditions[23] (more discussions in supplementary information section 8).

We find that the angular dependent intensity for $A_1$ modes varied in periods of 180° and 90° in parallel-polarized configuration; while those for $A_2$ phonon modes varied in a period of 90°. In particular, the $A_1$ modes having a period of 180° could be used to determine the crystallographic orientation of the TaIrTe$_4$ lattice, since their intensity maxima are exactly along the *a*- or *b*- axis of the crystals. Whether the intensity maxima pointing to the *a*- or *b*- axis of the crystals depends on the relative magnitude of matrix elements in $R_{A_1}$ for the specific phonon modes. Here we use the Raman mode at $100\ cm^{-1}$ and $148\ cm^{-1}$ in parallel-polarized configurations to identify the zigzag direction of the crystal (i.e. the direction of Ta-Ir chains). For these two modes, if *d* > *f*, then the intensity maxima is along the *a*- axis; if *d* < *f*, then the intensity maxima is along the *b*- axis (see supplementary information section 5 for detailed analysis). We experimentally determined that $d > f$ in TaIrTe$_4$ by performing polarized Raman spectroscopy and HR-TEM on the same flake. With this information, only polarized Raman spectroscopy is needed to rapidly and nondestructively identify the crystal orientation of TaIrTe$_4$ in future experiments. The recipe could be easily extended to other type-II Weyl semimetal candidates with in-plane anisotropy[20, 24].

**Electrical in-plane anisotropy**

According to DFT calculations, the density of states of TaIrTe$_4$ at the Fermi level is very high[7]. Experimentally we measured a Hall density of ~$1.2 \times 10^{27}$ m$^{-3}$. In view of the large carrier density, one can consider TaIrTe$_4$ to be one of the good metals, which usually have isotropic conductivity[25]. To our surprise, thin TaIrTe$_4$ samples exhibit extraordinarily strong electrical anisotropy, which is hitherto unprecedented among such good metals as copper and silver.

In order to study the electrical anisotropy of TaIrTe$_4$, we fabricated 12 electrodes (5 nm Cr/ 50 nm Au) on the same flake spaced at an angle of 30° along the directions as shown in **Figure 5**a and 0° is aligned roughly with the *a*-axis according to quantitative analysis based on the polarized Raman spectra. DC conductance was measured between each pair of diagonal contacts at different temperatures and the results are plotted in Figure 5b in polar coordinates. For an anisotropic material, the directional dependence of the low field conductivity can be described by the equation[26]:

$$\sigma_\theta = \sigma_x \cos^2(\theta - \phi) + \sigma_y \sin^2(\theta - \phi) \qquad (5)$$

$\sigma_x$ and $\sigma_y$ refer to the conductivity along the (100) and (010) directions, respectively. $\theta$ is the angle of the applied current with respect to the 0° reference direction, along which both the electric field is applied and the conductance is measured; $\phi$ is the angle between the x-direction and the 0° reference. Equation (5) fits very well to the measured data (solid curves in Figure 5b). From the fitting, we obtained a ratio of $\sigma_x/\sigma_y$ to be from 1.7 to 2.0 for temperature ranging from 300K to 10K (Figure 5c). It is found that the *a*- and *b* axes of the TaIrTe$_4$ thin film could be independently determined using the angle-resolved DC conductance measurement, with the maximum conductivity along the *a*-axis. In fact, the difference between the crystal orientations determined from the conductance measurement and from the polarized Raman spectroscopy is less than 5.6%. Another TaIrTe$_4$ device was measured at 300K and presented similar anisotropic behavior (more details in supplementary **Figure S6**).

To further understand the transport properties, we performed temperature-dependent Hall measurement along the *a*- and *b*- axes using the same device shown in Figure 5a. We attempted to obtain the ratio of Hall mobility ($\mu$) and carrier concentrations ($n$) along the two directions. Since Hall resistivity can be expressed as $R_{xy} = -\frac{B}{ne}$, we can obtain the carrier concentrations from the slopes of the $R_{xy}$ vs $B$ curves taken at different directions. To extract the slope along the *a* direction, a constant current *I* flows between leads 1 and 7 and the voltage *V* is measured between leads 4 and 10 when sweeping the magnetic field. Similarly, the slope of $R_{xy}$ vs $B$ along the *b* direction was obtained. We then use the formula $\sigma = ne\mu$ to get the ratio of the Hall mobilities. Figure 5c shows the temperature-dependent carrier concentration along *a*- and *b* axes ($n_a$ and $n_b$) and the inset shows $n_a/n_b$ and $\mu_a/\mu_b$ versus temperature. While the carrier concentrations remain the same along the two directions, a significant difference arises from the Hall mobility. Since the anisotropy in mobility decreases as temperature increases, we can exclude phonon scattering as the origin of the electrical anisotropy. A likely explanation of the anisotropy is the different effective masses along the two principle axes of the TaIrTe$_4$ crystals[27]. Such strong electrical anisotropy, together with the high carrier density and the environmental stability of the material (see supplementary information section 5 for stability study of TaIrTe$_4$ thin flakes), may enable new architectures in electrical interconnects in future integrated circuits[28].

**Summary**

In summary, we provide the first definitive evidence for the absence of inversion symmetry of the TaIrTe$_4$ crystals, by combining linearly polarized Raman spectroscopy and first-principle calculations. Our result shows that TaIrTe$_4$ could indeed be a type-II Weyl semimetal. We also demonstrate an efficient and nondestructive recipe to determine the exact crystallographic orientation of TaIrTe$_4$ crystals and other anisotropic type-II Weyl semimetal candidates. The in-plane electrical anisotropy of TaIrTe$_4$ thin flakes is found to be the strongest among materials with comparably high carrier density, which may enable new architectures in electrical interconnects in future integrated circuits.

**Methods**

**The growth of TaIrTe$_4$ single crystal.** All the used elements were stored and acquired in argon-filled glovebox with moisture and oxygen levels less than 0.1 ppm, and all manipulations were carried out in the glovebox. TaIrTe$_4$ single crystals were synthesized by solid state reaction with the help of Te flux. The elements of Ta powder (99.99 %), Ir powder (99.999 %), and Te lump (99.999 %) with an atomic ratio of Ta/Ir/Te = 1 : 1 : 12, purchased from Sigma-Aldrich (Singapore), were loaded in a quartz tube and then flame-sealed under high-vacuum of $10^{-6}$ torr. The quartz tube was placed in a tube furnace, slowly heated up to 1000 ℃ and held for 100 h, and then allowed to cool to 600 ℃ at a rate of 0.8 ℃/h, followed by a cool down to room temperature. The shiny, needle-shaped TaIrTe$_4$ single crystals can be obtained from the product.

**Raman spectroscopy.** The polarized Raman spectra were carried on an HR 800 (Jobin Yvon Horiba), with a 632.8 nm laser. A polarizer was used to select the incident polarization and was fixed during all the measurements. Another adjustable polarizer was added before the spectrometer to form the parrellel polarized and the cross polarized configurations. Raman measurements were carried out in ambient conditions, since the TaIrTe$_4$ flakes are fairly air stable (more details in supplementary **Figure S7**)[29].

**First-principles calculations.** The *ab initio* calculations are carried out using the Vienna *ab initio* simulation package (VASP)[30] with the local density approximation (LDA)[31] and the projector augmented wave potentials (PAW) [32]. The kinetic energy cutoff is fixed to 400 eV, and the *k*-point mesh is taken as $8 \times 2 \times 2$. The coordinates and the cell shape have been fully relaxed until the forces acting on the atoms are all smaller than $10^{-4}$ eV/Å, The small displacement method implemented in phonopy package[33] was used to get the phonon frequencies and vibration modes at the Γ point.

**Device fabrication and characterization.** Standard electron-beam lithography technique is used to pattern electrodes, consisting of 5 nm Cr and 50 nm Au, on the TaIrTe$_4$ samples to form devices. Electrical anisotropy measurements were carried out in a homemade low temperature ultra-high vacuum system with standard lock-in technique.

**Acknowledgement**
This project has been supported by the National Basic Research Program of China (973 Grant Nos. 2013CB921900, 2014CB920900, 2016YFA0301004), and the National Natural Science Foundation of China (NSFC Grant Nos. 11374021, 11774010, 11725415, 21573004) (Y.-N. Liu, Q.-Q. Gu, S.-M. Q, N. Zhang, Y.-N. Zhang, L.-M. Tong, H.-L. Peng, J. Feng and J.-H. Chen). Y. Peng and Z. Liu acknowledge support from Singapore National Research Foundation (NRF RF Award No. NRF-RF2013-08), MOE Tier 2 grant MOE2016-T2-1-131 (S).

**Author Contributions**
Y.N.L. and J.-H.C. conceived the experiment. Y.N.L. exfoliated the TaIrTe$_4$ thin flakes, fabricated the devices, completed AFM, Raman and transport measurements and analyzed the data. N.Z. helped with the Raman measurements under the supervision of L.M.T.. X.M.M and R.Z. performed TEM measurements. Q.Q.G. and J.F. performed the first principle calculations. Y.P. and Z.L. provided high quality bulk TaIrTe$_4$ crystals. Y.N.L., Q.Q.G., S.M.Q., Y.N.Z., J.F., Z.L. and J.-H.C. discussed the results and analyzed the data. Y.N.L., Q.Q.G., S.M.Q., J.F., Z.L. and J.-H.C. wrote the manuscript and all authors commented on it.

# REFERENCES

[1] A. A. Burkov, L. Balents, *Phys. Rev. Lett.* **2011**, 107.

[2] X. G. Wan, A. M. Turner, A. Vishwanath, S. Y. Savrasov, *Phys. Rev. B* **2011**, 83.

[3] M. N. X. Ali, J.; Flynn, S.; Tao, J.; Gibson, Q.D.; Schoop, L.M.; Liang, T.; Haldolaarachchige, N.; Hirschberger, M.; Ong, N.P.; Cava, R.J., *Nature* **2014**, 514, 205; X. Huang, L. Zhao, Y. Long, P. Wang, D. Chen, Z. Yang, H. Liang, M. Xue, H. Weng, Z. Fang, X. Dai, G. Chen, *Phys. Rev. X* **2015**, 5; B. Q. Lv, H. M. Weng, B. B. Fu, X. P. Wang, H. Miao, J. Ma, P. Richard, X. C. Huang, L. X. Zhao, G. F. Chen, Z. Fang, X. Dai, T. Qian, H. Ding, *Phys. Rev. X* **2015**, 5; H. Weng, C. Fang, Z. Fang, B. A. Bernevig, X. Dai, *Phys. Rev. X* **2015**, 5; S. Y. Xu, I. Belopolski, N. Alidoust, M. Neupane, G. Bian, C. L. Zhang, R. Sankar, G. Q. Chang, Z. J. Yuan, C. C. Lee, S. M. Huang, H. Zheng, J. Ma, D. S. Sanchez, B. K. Wang, A. Bansil, F. C. Chou, P. P. Shibayev, H. Lin, S. Jia, M. Z. Hasan, *Science* **2015**, 349, 613; S. M. Huang, S. Y. Xu, I. Belopolski, C. C. Lee, G. Q. Chang, B. K. Wang, N. Alidoust, G. Bian, M. Neupane, C. L. Zhang, S. Jia, A. Bansil, H. Lin, M. Z. Hasan, *Nat. Commun.* **2015**, 6.

[4] H. Weyl, *Z. Phys.* **1929**, 56, 330.

[5] A. A. Soluyanov, D. Gresch, Z. J. Wang, Q. S. Wu, M. Troyer, X. Dai, B. A. Bernevig, *Nature* **2015**, 527, 495.

[6] S. Murakami, *New Journal of Physics* **2007**, 9; P. Hosur, X. L. Qi, *C. R. Phys.* **2013**, 14, 857; A. M. V. Turner , A., *arXiv:1301.0330* **2013**.

[7] K. Koepernik, D. Kasinathan, D. V. Efremov, S. Khim, S. Borisenko, B. Buchner, J. van den Brink, *Phys. Rev. B* **2016**, 93.

[8] G. Xu, H. M. Weng, Z. J. Wang, X. Dai, Z. Fang, *Phys. Rev. Lett.* **2011**, 107.

[9] S. Y. Xu, N. Alidoust, G. Q. Chang, H. Lu, B. Singh, I. Belopolski, D. S. Sanchez, X. Zhang, G. Bian, H. Zheng, M. A. Husanu, Y. Bian, S. M. Huang, C. H. Hsu, T. R. Chang, H. T. Jeng, A. Bansil, T. Neupert, V. N. Strocov, H. Lin, S. A. Jia, M. Z. Hasan, *Science Advances* **2017**, 3; S. Borisenko, D. Evtushinsky, Q. Gibson, A. Yaresko, T. Kim, M. N. Ali, B. Buechner, M. Hoesch, R. J. Cava, *Physics* **2015**; Y. Sun, S. C. Wu, M. N. Ali, C. Felser, B. H. Yan, *Phys. Rev. B* **2015**, 92; G. Q. Chang, S. Y. Xu, D. S. Sanchez, S. M. Huang, C. C. Lee, T. R. Chang, G. Bian, H. Zheng, I. Belopolski, N. Alidoust, H. T. Jeng, A. Bansil, H. Lin, M. Z. Hasan, *Science Advances* **2016**, 2.

[10] Z. J. Wang, D. Gresch, A. A. Soluyanov, W. W. Xie, S. Kushwaha, X. Dai, M. Troyer, R. J. Cava, B. A. Bernevig, *Phys. Rev. Lett.* **2016**, 117.

[11] A. Mar, S. Jobic, J. A. Ibers, *J. Am. Chem. Soc.* **1992**, 114, 8963.

[12] C. J. Bradley, A. P. Cracknell, *The Mathematical Theory of Symmetry in Solids*, *Clarendon Press*, **1972**; G. F. Koster, *Solid State Physics: Advances in Research and Applications* **1957**, 5, 173.

[13] I. Y. Belopolski, P.; Zahid Hasa, M., *arXiv:1610.02013.* **2016**.

[14] E. Haubold, K. Koepernik, D. Efremov, S. Khim, A. Fedorov, Y. Kushnirenko, J. van den Brink, S. Wurmehl, B. Buchner, T. K. Kim, M. Hoesch, K. Sumida, K. Taguchi, T. Yoshikawa, A. Kimura, T. Okuda, S. V. Borisenko, *Phys. Rev. B* **2017**, 95.

[15] I. Belopolski, D. S. Sanchez, Y. Ishida, X. C. Pan, P. Yu, S. Y. Xu, G. Q. Chang, T. R. Chang, H. Zheng, N. Alidoust, G. Bian, M. Neupane, S. M. Huang, C. C. Lee, Y. Song, H. J. Bu, G. H. Wang, S. S. Li, G. Eda, H. T. Jeng, T. Kondo, H. Lin, Z. Liu, F. Q. Song, S. Shin, M. Z. Hasan, *Nat. Commun.* **2016**, 7; Y. J. Wang, E. F. Liu, H. M. Liu, Y. M. Pan, L. Q. Zhang, J. W. Zeng, Y. J. Fu, M. Wang, K. Xu, Z.

Huang, Z. L. Wang, H. Z. Lu, D. Y. Xing, B. G. Wang, X. G. Wan, F. Miao, *Nat. Commun.* **2016**, 7.

[16] Y. Y. Wang, Z. H. Ni, Z. X. Shen, H. M. Wang, Y. H. Wu, *Appl. Phys. Lett.* **2008**, 92; X. Ling, J. Zhang, *J. Phys. Chem. C* **2011**, 115, 2835.

[17] Z. B. Yang, W. J. Jie, C. H. Mak, S. H. Lin, H. H. Lin, X. F. Yang, F. Yan, S. P. Lau, J. H. Hao, *ACS Nano* **2017**, 11, 4225; I. Stenger, L. Schue, M. Boukhicha, B. Berini, B. Placais, A. Loiseau, J. Barjon, *2D Mater.* **2017**, 4.

[18] R. Loudon, *Adv. Phys.* **1964**, 13, 423.

[19] W. D. Kong, S. F. Wu, P. Richard, C. S. Lian, J. T. Wang, C. L. Yang, Y. G. Shi, H. Ding, *Appl. Phys. Lett.* **2015**, 106; Y. C. Jiang, J. Gao, L. Wang, *Sci. Rep.* **2016**, 6.

[20] Q. J. Song, X. C. Pan, H. F. Wang, K. Zhang, Q. H. Tan, P. Li, Y. Wan, Y. L. Wang, X. L. Xu, M. L. Lin, X. G. Wan, F. Q. Song, L. Dai, *Sci. Rep.* **2016**, 6.

[21] M. Kim, S. Han, J. H. Kim, J. U. Lee, Z. Lee, H. Cheong, *2D Mater.* **2016**, 3.

[22] N. N. Mao, J. X. Wu, B. W. Han, J. J. Lin, L. M. Tong, J. Zhang, *Small* **2016**, 12, 2627; S. Zhang, N. Mao, N. Zhang, J. Wu, L. Tong, J. Zhang, *ACS Nano* **2017**.

[23] S. Y. Chen, C. H. Naylor, T. Goldstein, A. T. C. Johnson, J. Yan, *ACS Nano* **2017**, 11, 814.

[24] F. N. Xia, H. Wang, Y. C. Jia, *Nat. Commun.* **2014**, 5; R. He, J. A. Yan, Z. Y. Yin, Z. P. Ye, G. H. Ye, J. Cheng, J. Li, C. H. Lui, *Nano Lett.* **2016**, 16, 1404; K. N. Zhang, C. H. Bao, Q. Q. Gu, X. Ren, H. X. Zhang, K. Deng, Y. Wu, Y. Li, J. Feng, S. Y. Zhou, *Nat. Commun.* **2016**, 7; X. M. Wang, A. M. Jones, K. L. Seyler, V. Tran, Y. C. Jia, H. Zhao, H. Wang, L. Yang, X. D. Xu, F. N. Xia, *Nat. Nanotechnol.* **2015**, 10, 517; S. Y. Chen, T. Goldstein, D. Venkatararnan, A. Ramasubramaniam, J. Yan, *Nano Lett.* **2016**, 16, 5852.

[25] N. W. A. N. D. Mermin, *Solid State Physics*, *Harcourt College Publishers*, **1976**; C. Kittel, *Introduction to Solid State Physics*, *John Wiley & Sons, Inc.*, **2005**.

[26] G. Qiu, Y. C. Du, A. Charnas, H. Zhou, S. Y. Jin, Z. Luo, D. Y. Zemlyanov, X. F. Xu, G. J. Cheng, P. D. D. Ye, *Nano Lett.* **2016**, 16, 7364; W. Yu, Y. Jiang, J. Yang, Z. L. Dun, H. D. Zhou, Z. Jiang, P. Lu, W. Pan, *Sci. Rep.* **2016**, 6.

[27] D. Chen, L. X. Zhao, J. B. He, H. Liang, S. Zhang, C. H. Li, L. Shan, S. C. Wang, Z. A. Ren, C. Ren, G. F. Chen, *Phys. Rev. B* **2016**, 94; S. H. Khim, K. Koepernik, D. V. Efremov, J. Klotz, T. Forster, J. Wosnitza, M. I. Sturza, S. Wurmehl, C. Hess, J. van den Brink, B. Buchner, *Phys. Rev. B* **2016**, 94; X. L. Xu, Q. J. Song, H. F. Wang, P. Li, K. Zhang, Y. L. Wang, K. Yuan, Z. C. Yang, Y. Ye, L. Dai, *ACS Appl. Mater. Interfaces* **2017**, 9, 12601.

[28] P. S. Bhardwaj, P.; Uma Sathyakam P.; Karthikeyan A. , *Imperial Journal of Interdisciplinary Research* **2016**, 2; A. Hosseini, V. Shabro, *Microelectron. Eng.* **2010**, 87, 1955.

[29] H. M. Oh, G. H. Han, H. Kim, J. J. Bae, M. S. Jeong, Y. H. Lee, *ACS Nano* **2016**, 10, 5230.

[30] G. Kresse, J. Furthmuller, *Phys. Rev. B* **1996**, 54, 11169.

[31] J. P. Perdew, A. Zunger, *Phys. Rev. B* **1981**, 23, 5048.

[32] G. Kresse, D. Joubert, *Phys. Rev. B* **1999**, 59, 1758.

[33] A. Togo, I. Tanaka, *Scr. Mater.* **2015**, 108, 1.

**Figures**

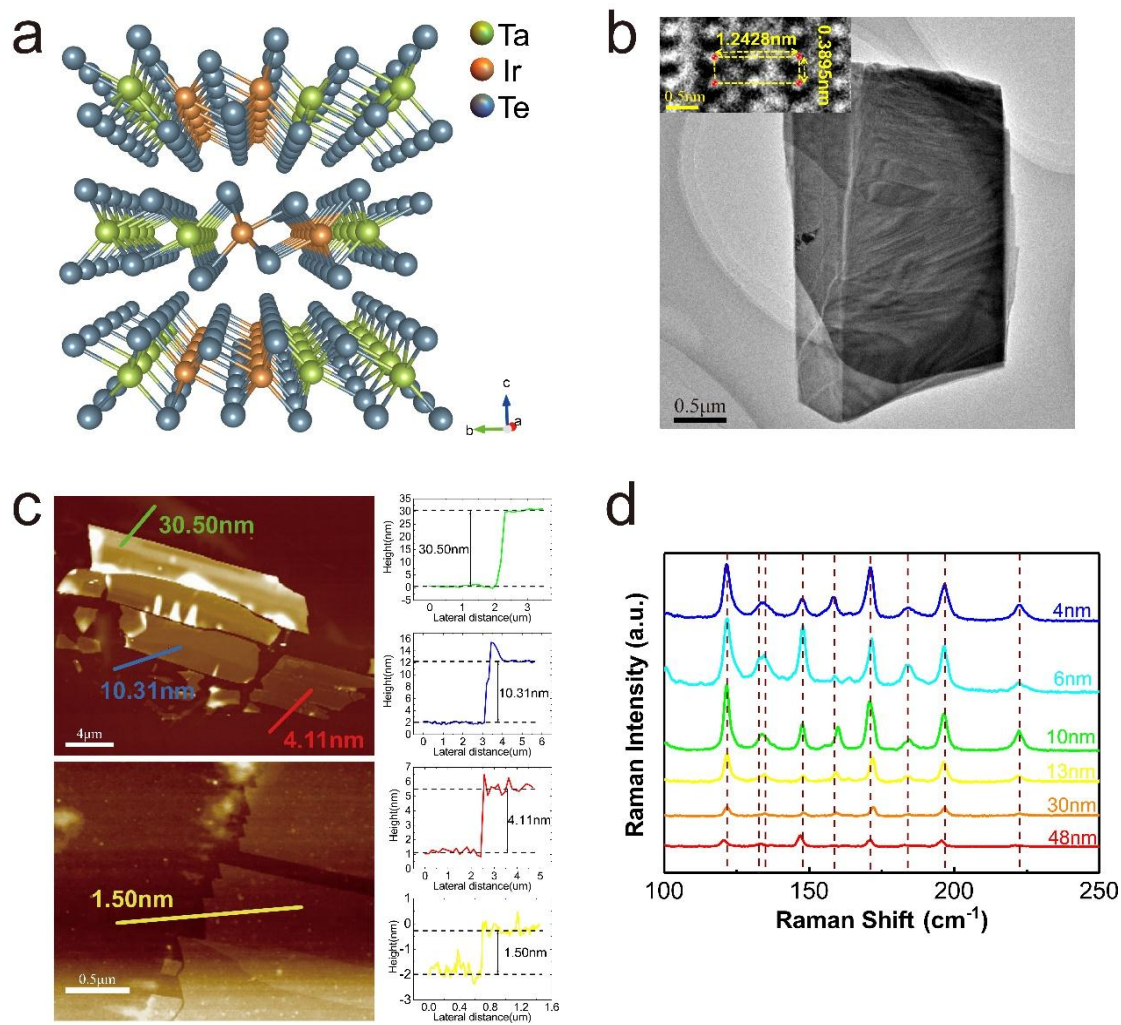

**Figure 1.** (a) Crystal structure of TaIrTe$_4$. (b) TEM image of a TaIrTe$_4$ thin flake. Inset: the corresponding HR-TEM image of the TaIrTe$_4$ thin flake. (c) AFM images of TaIrTe$_4$ flakes with various thickness. (d) Unpolarized Raman spectra measured at TaIrTe$_4$ flakes of different thicknesses.

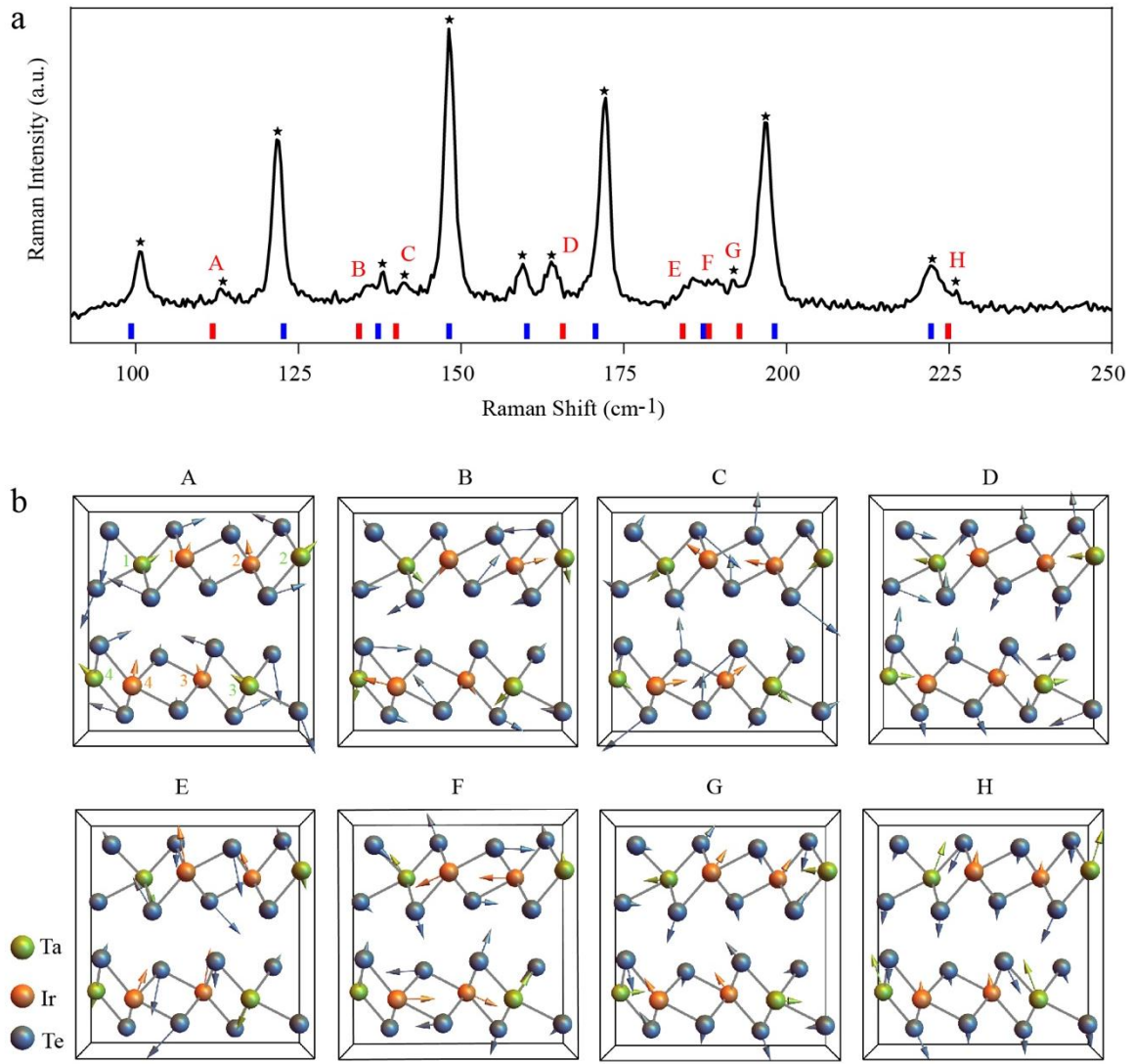

**Figure 2.** Polarized Raman spectra and phonon vibrational patterns. (a) Polarized Raman spectra (solid black curve) with the parallel polarization configuration along *a*-axis and the respective calculated frequencies (vertical bars) of the $A_1$ representation phonons. The black stars show the central frequencies of the measured Raman peaks. (b) Calculated phonon vibrational patterns for the Raman modes that are highlighted with red bars in (a). The green numbers 1 to 4 in vibrational pattern A denote the four Ta atoms and the orange number 1 to 4 denote the four Ir atoms in the unit cell, in which 1 and 2 denotes the inversion equivalent atoms in the hypothetical inversion symmetric structure.

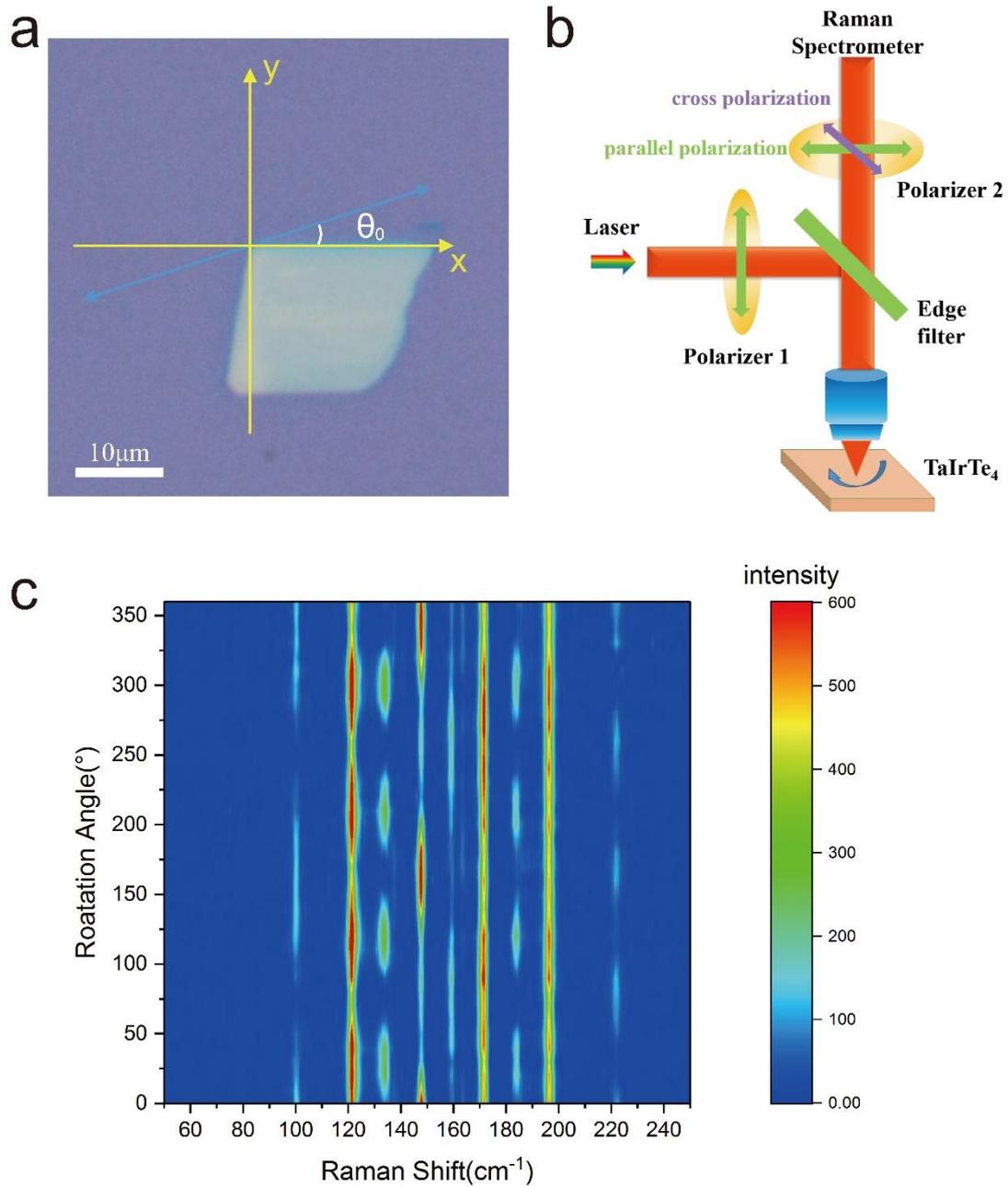

**Figure 3.** (a) An optical image of a TaIrTe$_4$ flake with a thickness of around 12 nm. The blue double-headed arrow indicates the polarization of the incident light. (b) Schematic diagram of the angle-resolved polarized Raman spectroscopy of TaIrTe$_4$ samples. (c) Angular dependence of the intensity of the Raman spectra for a freshly exfoliated flake measured in parallel-polarized configuration.

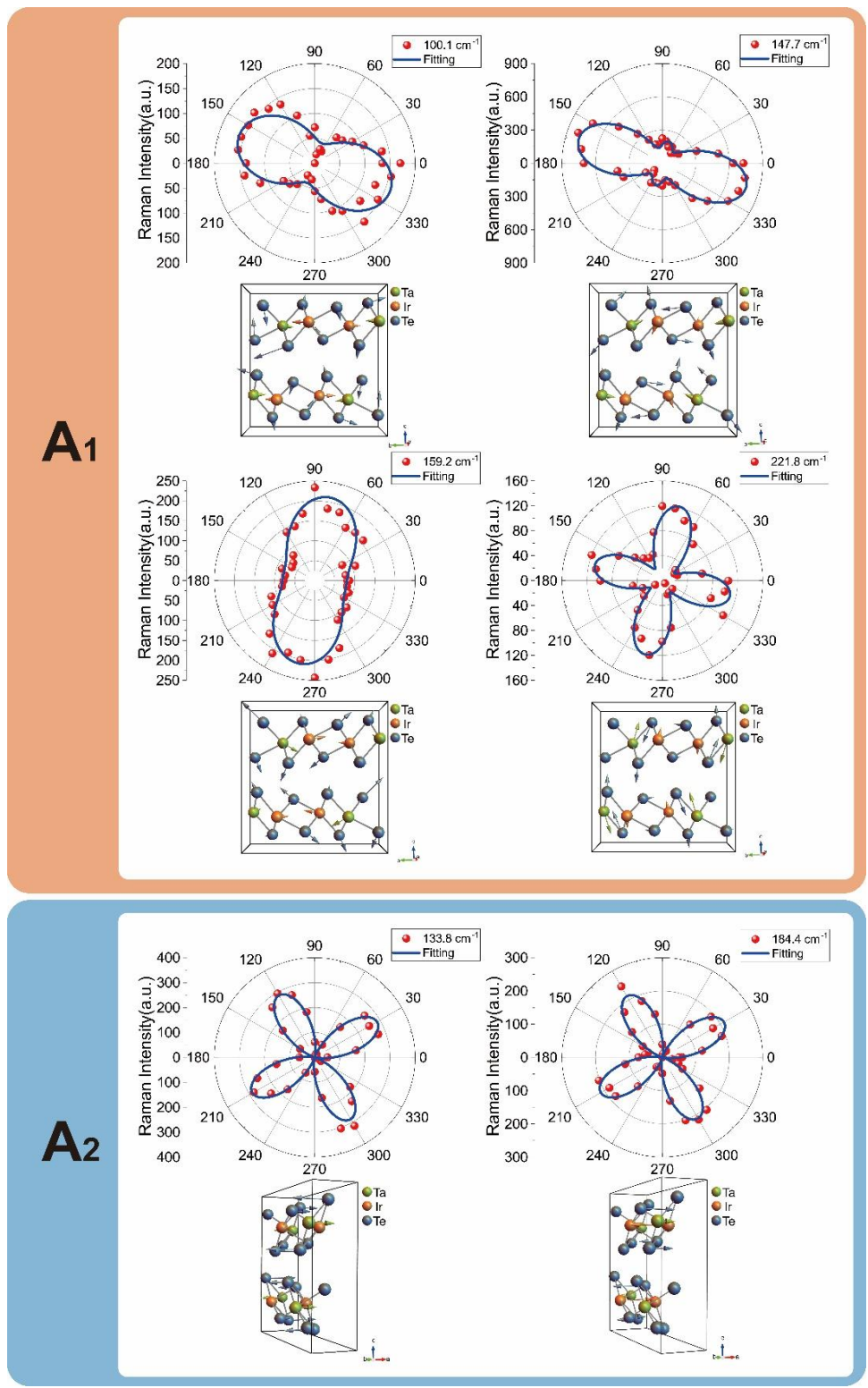

**Figure 4.** Raman intensities of six modes as a function of the sample rotation angle and the corresponding phonon modes in atomic view. The scattered dots are experimental data, and the solid lines are fitting curves.

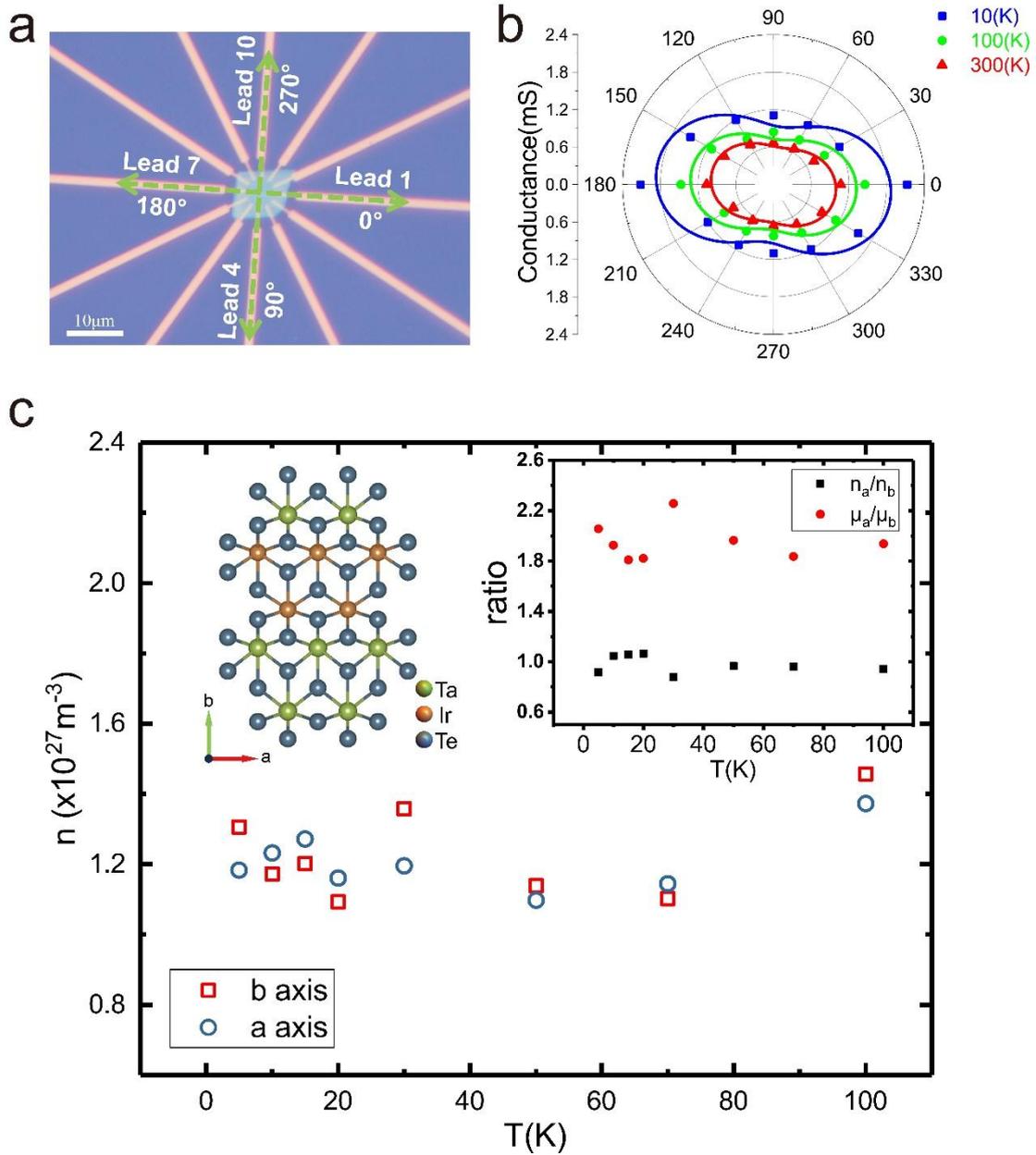

**Figure 5.** Angle-resolved DC conductance measurements of TaIrTe$_4$ thin flakes. (a) An optical image of the TaIrTe$_4$ thin flake device. (b) Angle-dependent DC conductance at different temperatures. (c) Temperature-dependent carrier concentration measured along *a*- and *b*- axes. Upper left inset: atomic top view of the *ab* plane of the TaIrTe$_4$ crystal; upper right Inset: the ratio of Hall mobilities along *a*- and *b*- axes at different temperatures.